\documentclass[runningheads]{llncs}
\usepackage{graphicx}
\usepackage{algorithm,algorithmic}
\usepackage[figuresright]{rotating}
\usepackage{amsmath}
\begin{document}
\title{New Competitive Analysis Results of
Online List Scheduling Algorithm\thanks{Supported by Veer Surendra Sai  University of Technology}}
\author{Rakesh Mohanty\inst{1} \and
Debasis Dwibedy\inst{1} \and
Shreyaa Swagatika Sahoo\inst{1}}
\authorrunning{R. Mohanty et al.}
\institute{Veer Surendra Sai University of Technology, Burla,  Odisha 768018, India 
%\email{rakesh.iitmphd@gmail.com}
\email{\{rakesh.iitmphd, debasis.dwibedy, shryaa.swagatika\}@gmail.com}}
\maketitle              % typeset the header of the contribution
\begin{abstract}
Online algorithm has been an emerging area of interest for researchers in various domains of computer science. The online $m$-machine list scheduling problem introduced by Graham has gained theoretical as well as practical significance in the development of competitive analysis as a performance measure for online algorithms. In this paper, we study and explore the performance of Graham's online \textit{list scheduling algorithm(LSA)} for independent jobs. In the literature, \textit{LSA} has already been proved to be $2-\frac{1}{m}$ competitive, where $m$ is the number of machines. We present two new upper bound results on competitive analysis of \textit{LSA}. We obtain upper bounds on the competitive ratio of $2-\frac{2}{m}$ and $2-\frac{m^2-m+1}{m^2}$ respectively for practically significant two special classes of input job sequences. Our analytical results can motivate the practitioners to design improved competitive online algorithms for the  $m$-machine list scheduling problem by characterization of real life input sequences.
\keywords{Online Algorithm \and Competitive Analysis \and Scheduling \and Identical Machines \and Non-preemptive \and Makespan.}
\end{abstract}
\subsection{Online Algorithm} \label{subsec: Online Algorithm}
%The algorithm in which input information is partially available prior to the processing of the inputs is known as \textit{online algorithm} [2].\\
An \textit{online algorithm} receives and processes inputs one by one in order [1, 2]. Each input is processed immediately upon its availability with no knowledge on the successive inputs. Since, the algorithm has no prior idea about the entire sequence of inputs, it is constrained to make irrevocable decisions on the fly. Here, a sequence of outputs are produced by considering each time the past outputs and the current input. Suppose, we have a sequence of inputs $ I=\left<i_1, i_2,.........i_n \right>$ of finite size $n$. The inputs are available to the online algorithm one at a time so that at any given time $t$ an input instance $i_t$ is processed with no clue on the future inputs $i_{t'}$, where $t'>t$. \\
Interactive computing is indispensable in various domains such as computers, networks, transport, medical, agriculture, production and industrial management [2]. Online algorithms can be extremely useful for interactive computing. The requests arrive one by one to the interactive system and each request demands an immediate response. Here, the system runs an online algorithm that reacts to the current request according to the desired objective and without knowledge of the entire request sequence. Therefore, design and analysis of online algorithms have gained a serious research interest and practical significance.
\subsection{Competitive Analysis} \label{subsec: Competitive Analysis} 
Competitive analysis [4] provides a theoretical framework to measure the performance of an online algorithm. Here, the performance of an online algorithm is compared with its corresponding optimum offline algorithm which knows all information about the inputs a priori and processes them efficiently by incurring smallest cost. Let us consider $ALG(I)$ be the cost incurred by an online algorithm $ALG$ for any input sequence $I$ and $OPT(I)$ be the  optimum cost obtained by the optimum offline algorithm $OPT$ for $I$. We now define $ALG$ to be $k$-competitive for a smallest $k\geq 1$, if $ALG(I) \leq k\cdot OPT(I)$ for all input sequences $I$. Here, $k$ is referred to as the competitive ratio. For a cost minimization problem, it is always desirable to obtain the competitive ratio which is closer to $1$.
\subsection{Online List Scheduling} \label{subsec: Online Scheduling}
\textit{Online List scheduling}(\textit{LS}) [1] has been a well studied problem in theoretical computer science. Here we are given a finite number of jobs in  a list and $m$-machines($m \geq 2$). The output is the generation of a schedule which represents the assignments of all jobs over $m$ machines, where the completion time of the job schedule i.e. \textit{makespan} is the output parameter. The objective is to attain a minimum makespan subject to some non-trivial constraints. The constraints are- input jobs are revealed one by one. Each available job must be scheduled irrevocably as soon as it is given with no information about the successive jobs. The assumptions are that jobs are non-preemptive and independent.
\subsection{Practical and Research Motivation} \label{subsec: Practical and Research Motivation}
Online list scheduling finds applications in areas such as multiprocessor scheduling in the interactive time shared operating systems [2], routing of data packets on different links with balancing the loads of each link in the computer networks [6], data and information processing in the distributed computing systems [7], robot navigation and exploration [8].    \\
%\subsection{Research Motivation}
%\label{subsec:Research Motivation}
Online $m$-machine list scheduling for $m \geq 2$ has been proved to be NP-Complete by a polynomial time reduction from the classical Partition problem [9]. The real challenge for designing of near optimal online scheduling algorithm arises due to the unavailability of required information on the entire job sequence prior to their processing. Basically, an online list scheduling algorithm is influenced by the sequence of arrival of the input jobs and their processing times. According to our knowledge, there is no attempt in the literature to classify and characterize the input job sequences for online list scheduling  based on real world inputs. This motivates us to study and analyze the widely accepted and practically implemented online list scheduling algorithm \textit{LSA } by exploring and characterizing special classes of inputs. 
\subsection{Contributions} \label{subsec: Contributions} 
We characterize the performance of algorithm \textit{LSA} for online scheduling of independent jobs on $m$ identical parallel machines and present a simple proof for $2-\frac{1}{m}$ competitiveness. We analyze algorithm \textit{LSA} on special classes of job sequences and obtain two new upper bounds on the competitive ratio as $2-\frac{2}{m}$ and $2-\frac{m^2-m+1}{m^2}$ respectively.

\section{Preliminaries and Related Work}\label{sec:Background and Related Work}
Here, we present some basic terminologies and notations, which we will use through out the paper. We then highlight scholarly contributions related to online list scheduling setting. 
%In this paper, we study a realistic scenario of \textit{LS} in which jobs arrive one by one and we call the problem as online list scheduling or simply \textit{Online Scheduling}. \\
\subsection{Basic Teminologies and Notations}\label{sec:Basic Teminologies and Notations}
\begin{itemize}
\item
We specify each independent job and identical machine as $J_i$ and $M_j$ respectively, where $m$ machines are represented as $M_j(j=1......m)$ and $n$ jobs are represented as $J_i(i=1......n)$. \item Jobs are \textit{independent} in the sense that jobs can execute in overlapping time slots on different machines. \item  Machines are \textit{identical} in the sense that the processing time ($p_i$) of  any $J_i$ is equal for all machines. \item Sometime we refer processing time ($p_i$) of $J_i$ as \textit{size} of $J_i$. \item We represent $c_i$ as the completion time of any job $J_i$. \item We denote \textit{makespan} obtain by any online algorithm \textit{A} for input sequence \textit{I} as $C^*_{A}(I)$. We have $C^*_{A}(I)$=$max\{c_i|1 \leq i \leq n\}$. \item A machine is in idle state when it is not executing any job and we represent idle time of the machine as $\varphi$ in the timing diagram.  \item \textit{Load}($l_j$) of any machine $M_j$ is the sum of processing time of the jobs scheduled on $M_j$. Suppose, $n$ jobs are assigned to $M_j$, then $l_j$=$\sum_{i=1}^{n}{p_i}$. We may further define makespan as $C^*_{A}(I)$=$max\{l_i|1 \leq j \leq m\}$. \item  \textit{Non-preemptive} scheduling of the jobs means once a job $J_i$ with $p_i$ starts its processing on any $M_j$ at time $t$ then it continues with no interruption by taking all together $t+p_i$ time prior to its completion.
\end{itemize}
\subsection{Related Work}\label{sec:Related Work}
The $m$-machine \textit{LS} problem has been studied for various setups over the years, see surveys [10-13]. According to our knowledge, the first online scheduling algorithm for multiprocessor systems was proposed by Graham in 1966 popularly known as list scheduling algorithm(\textit{LSA}) [1]. He considered the non-preemptive scheduling of a list of jobs on identical parallel machines. The goal was to obtain minimum makespan. Algorithm \textit{LSA} schedules a newly available job to the most lightly loaded machine. The performance of \textit{LSA} was proved to be at most $2-\frac{1}{m}$ time worse than the optimum makespan for all job sequences.\\
%Although, there have been notable efforts to improve the performance of \textit{LSA} by designing different heuristics, no attempt has been made for improving the performance of \textit{LSA} by characterization of input job sequence.\\
Faigle et. al. [5] analyzed the performance of \textit{LSA} by considering a list of $3$ jobs with sizes ($1,1,2$) respectively and proved that \textit{LSA} is optimal for $m=2$. Similarly, for $m=3$, they considered $7$ jobs with sizes ($1,1,1,3,3,3,6$) respectively to represent the optimum competitiveness of $1.66$. They obtained \textit{lower bound(LB}) on the competitive ratio of $1.707$ for $m \geq 4$ by considering a list of $2m+1$ jobs, where $m$ jobs are of size $1$ unit each, $m$ jobs with size $1+\sqrt{2}$ unit each and a single job is of size $2(1+\sqrt{2})$ unit.\\
The first improvement over \textit{LSA} was provided by Galambos and Woeginger [14] and achieved competitiveness of ($2-\frac{1}{m}-\epsilon_{m}$), where $\epsilon_{m} > 0$. Bartal et. al. [15] obtained the upper bound(UB) on the competitive ratio of $1.986$ for a general case of $m$. For $m=3$, they proved LB of $1.4$ by considering $7$ jobs each with size ($1,1,1,2,1,3,5$) unit respectively.
% Karger et. al. [16] improved the upper bound to $1.945$ for large $m$.\\
Bartal et. al. [17] obtained a better \textit{LB} of $1.837$ for $m \geq 3454$ by examining the job sequence consisting of $4m+1$ jobs, where $m$ jobs each with processing time of $\frac{1}{x+1}$ unit, $m$ jobs with processing time of $\frac{x}{x+1}$ unit each, $m$ jobs are of size $x$ unit each, $\lfloor\frac{m}{2}\rfloor$ jobs are of size $y$ unit each, $\lfloor\frac{m}{3}\rfloor-2$ jobs each with size $z$ unit, ($m+3-\lfloor\frac{m}{2}\rfloor-\lfloor\frac{m}{3}\rfloor$) jobs are of size $2y$ unit each, where $x,y,z$ are positive real values. We now present the summary of all important results for deterministic online scheduling algorithms for identical parallel machines in table \ref{tab:summary of Important Results}. 
\begin{table}[htbp]
%\centering
\caption{Summary of Important Results}
\begin{tabular} {|c|p{5.5cm}|}
\hline
\textbf{Year and Author(s)}  & \textbf{Competitive Ratio(s)} \\
\hline 
1966, Graham [1] &  $2-(\frac{1}{m})$ for all $m$.\\
\hline
1991, Galambos and Woeginger [14] &  $2-(\frac{1}{m} - \epsilon_m)$ for $m \geq 4$.\\
\hline
1992, Bartal et al. [15] &  $1.986$(UB) for all $m$, $1.4$(LB)  for $m=3$.\\
\hline
1996, Karger et al. [16] &  $1.945$, for $m \geq 8$.\\
\hline 
1994, Bartal et. al. [17] & $1.837$(LB) for $m\geq3454$ \\
\hline
1994, Chen et.al. [18] & $1.7310$(LB) for $m=4$ and $1.8319$(LB) for $m > 4$. \\
\hline
1999, Albers [19] &  $1.923$, for $m \geq 2$.\\
\hline 
2000, Fleicher and Wahl [20] & $1.9201$(UB)\\
\hline
2001, Rudin III [21] & $1.88$(LB) for all $m$.\\
\hline
2003, Rudin III and Chnadrasekharan [22] & $1.732$(LB) for $m=4$\\
\hline
2008, Englert et.al. [23] & $1.4659$ for $2 \leq m \leq 30 $\\
\hline 
\end{tabular}
\label{tab:summary of Important Results}
\end{table}
\subsection{Graham's  Online List Scheduling Algorithm}\label{sec:Graham's  Online List Scheduling Algorithm}
Here, we present the descriptions of algorithm \textit{LSA} [1] for independent jobs and provide proof sketch to show its competitiveness results as follows.
%In this study, we are interested in the online scheduling problem that resembles with Graham's list scheduling for independent jobs. Therefore, in the next section we present Graham's \textit{LSA} for independent jobs. 

%\subsection{LSA and its Competiveness} \label{subsec:List LSA and its Competiveness}
%\textbf{LSA:} \\\\
\begin{algorithm}
\caption{LSA}
\begin{algorithmic}
  \scriptsize
\STATE Initially, i=1, $l_1=l_2=.............l_m=0$\\
\STATE WHILE a new job $J_{i}$ arrives DO\\
\STATE \hspace*{0.2cm} BEGIN\\
\STATE \hspace*{0.5cm} Calculate current load for each machine $M_j$. \\
\STATE \hspace*{0.5cm} Number the machines in non-decreasing order of their loads
Such that \hspace*{0.5cm} $l_1 \leq  l_2  \leq ......\leq l_m$. \\
\STATE \hspace*{0.5cm} Assign $J_{i}$ to $M_1$.\\
\STATE \hspace*{0.5cm} $l_1=l_1+p_{i}$ 
\STATE \hspace*{0.5cm} $i=i+1$.\\
\STATE \hspace*{0.2cm} END\\
\STATE Return \hspace*{0.3cm} $l_j=max\{l_j|j=1, 2,......m\}$
\end{algorithmic}
\end{algorithm} 

\textbf{Theorem 1}. Algorithm LSA is ($2-\frac{1}{m}$)-competitive for $m\geq 2$.
\\
\textit{Proof:} Let us consider a list of $n$ jobs($J_1........J_n$). Each job is available to \textit{LSA} one by one. The processing time $p_i > 0$ for $1 \leq i \leq n$. Initially, $m$ machines($M_1,.....M_m$)  are available with loads $l_1=l_2=.....=l_m=0$. Let the size of the largest job $J_k$ is $p_k$, where $p_k$ =$ \max\{p_i|1 \leq i \leq n\}$. We denote the optimal makespan as $C^*_{OPT}(I)$ and makespan obtained by algorithm \textit{LSA} as $C^*_{LSA}(I)$ for all input sequences \textit{I}. As per the description of \textit{LSA}, the scheduling decision time($T$) is constant for each input. Therefore, each time we ignore $T$, while calculating makespan.\\
\textit{Computation of OPT:} Optimum offline strategy  equally distributes the total load among all $m$-machines. So, the completion time of the job schedule is at least the average of total load incurred on $m$-machines. Therefore, we have\\ 
\hspace* {3.2cm} $C^*_{OPT}(I) \geq \frac{1}{ m}(\sum_{i=1}^{n}{p_i})$. \hspace* {4.6cm}(1)\\
Suppose \textit{OPT} schedules only $J_k$ on $M_1$ and assigns rest $n-1$ jobs on $m-1$ machines with equal load sharing among $m-1$ machines and   $\frac{1}{m-1}(\sum_{i=2}^{m}{l_i}) \leq l_1$,
% average load on $m-1$ machines is less than or equal to the size of the largest job, \\
 then we have \\
 \hspace* {4.2cm} $C^*_{OPT}(I) \geq p_k$. \hspace* {5.1cm}(2)\\\\
\textit{We now provide the computation for algorithm LSA :} Algorithm \textit{LSA} assigns a new job to the machine with least load to keep a balance in the load incurred on each machine. The worst scenario appears in this case when $J_k$ arrives as the $n^{th}$ job and prior to that the total load incurred by  $(n-1)$ jobs are equally shared among $m$-machines. So, we have $l_1 \leq \frac{1}{m} (\sum_{i=1}^{n-1}{p_i})$, this compels \textit{LSA} to schedule the $n^{th}$ job on $M_1$ i.e the least loaded machine. Therefore, we have \\
$C^*_{LSA}(I) \leq \frac{1}{m}(\sum_{i=1}^{n-1}{p_i})+p_k $ \\ $m.C^*_{LSA}(I) \leq \sum_{i=1}^{n-1}{p_i}+ m.(p_k) \leq \sum_{i=1}^{n-1}{p_i} + p_k + (m-1).p_k \leq \sum_{i=1}^{n}{p_i} + (m-1).C^*_{OPT}(I)$ \\ $C^*_{LSA}(I) \leq \frac{1}{m}(\sum_{i=1}^{n}{p_i})+(\frac{m-1}{m}).C^*_{OPT}(I) \leq  C^*_{OPT}(I) + (\frac{m-1}{m}).C^*_{OPT}(I) \leq C^*_{OPT}(I)(1+\frac{m-1}{m})$ \\ $\frac{C^*_{LSA}(I)}{C^*_{OPT}(I)} \leq \frac{m+m-1}{m} \leq \frac{2m-1}{m}$ \\ $ C^*_{LSA}(I) \leq (2-\frac{1}{m}){C^*_{OPT}(I)}$
\section{New Upper Bound Results on Competitiveness of Algorithm LSA}
\label{subsec:New Upper bound Results on Competitiveness of LSA}
We obtain improved competitive ratios of the online deterministic \textit{LSA} by considering two special classes of inputs. In this setting, the performance of \textit{LSA} is evaluated through the ratio between the makespan obtained by \textit{LSA} for worst sequence of input jobs arrival to the makespan obtained by \textit{OPT}. The special classes of input sequences are described as follows.      
\subsection{ Special Classes of Input Job Sequences} 
\label{subsec:Special Classes of Input Job Sequences}
\textit{Class-$1$(${S_1}$):} 
Here, we consider a list of $(m-1)^2 + 1$ jobs, where $(m-1)^2$ number of jobs are of size $1$ unit each and a single job is of size $m$ unit.\\\\
% Suppose, the jobs arrive one by one in the following order $\sigma_1= \left<J_1, J_2, .......J_{(m-1)^2}, J_{(m-1)^2+1}\right>$, where the jobs from $J_1$ to $J_{(m-1)^2}$ are of size $1$ unit each and the ${J_{(m-1)^2+1}}^{th}$ job is of size $m$ unit.
\textit{Class-$2$(${S}_2$):}  
Here, we consider a list of $m(m-1)+ 1$ jobs, where $m(m-1)$ number of jobs are of size $1$ unit each and a single job is of size $m^2$ unit.\\\\
\textbf{Theorem 2}. \textit{LSA is ($2-\frac{2}{m}$)-competitive for ${S}_{1}$, where $m\geq3$.}\\\\
\textit{Proof:} Let, $C^{*}_{OPT}({S}_{1})$ and $C^{*}_{LSA}({S}_{1})$ be the makespan obtained by \textit{OPT} and \textit{LSA } respectively for ${S}_{1}$. We ignore $T$, while scheduling each incoming job.\\\\
\textit{Computation of LSA:} The worst sequence for ${S}_{1}$ appears when the input jobs arrive in the  non-decreasing order of their processing time. So, in the worst case, jobs arrive one by one starting at time $t=0$ in the following order $\sigma_1= \left<J_1, J_2, .......J_{(m-1)^2}, J_{(m-1)^2+1}\right>$, where the jobs from $J_1$ to $J_{(m-1)^2}$ are of size $1$ unit each and the $({J_{(m-1)^2+1}})^{th}$ job is of size $m$ unit. \textit{LSA} schedules each job upon its availability and before the arrival of the next job. As we are ignoring $T$, so at time $t=0$, $m$ jobs are scheduled on $m$ machines in one slot to complete their processing at $t=1$. Therefore, the first $m^2-2m$ jobs finish at $t=m-2$. Now, at $t=(m-2)$, we are left with final two jobs of sizes $1$ and $m$ respectively and are allocated to machines $M_1$ and $M_2$. So, the last job finishes at $t=2m-2$.
Therefore, we have \\ 
\hspace* {3.2cm} $C^*_{LSA}({S}_{1}) \leq 2m-2$ \hspace*{5.2cm}(3)\\     
%We present the Gantt chart of $C^*_{LSA}({S}_{1})$ in figure \ref{fig:ganttchartoflsacase1.png}.\\
%\begin{figure}[h]
%centering
%\includegraphics[scale=0.55]{ganttchartoflsacase1.png}
%\caption{Gantt chart of LSA for class 1}
%\label{fig:ganttchartoflsacase1.png}  
%\end{figure} 

\textit{Computation of OPT:} Here, the optimum strategy schedules the jobs according to the non-increasing order of job's size. So, at time $t=0$, \textit{OPT} assigns the largest job with size $m$ unit to a machine along with  $m-1$ jobs of size $1$ unit each to rest $m-1$ machines. Subsequently, $(m-1)^2$ jobs are assigned and completed at $t=m-1$ and the last job finishes at $t=m$. 
Therefore, we have\\
\hspace* {3.2cm} $C^*_{OPT}({S}_{1}) \geq m$ \hspace*{6.1cm}(4)\\ 
%The Gantt chart of $C^*_{OPT}({S}_{1})$ is presented in figure \ref{fig:ganttchartofoptimalcase_1.png}.
%\begin{figure}[h]
%\centering
%\includegraphics[scale=0.54]{ganttchartofoptimalcase_1.png}
%\caption{Gantt chart of Optimal Scheduling for class 1}
%\label{fig:ganttchartofoptimalcase_1.png}  
%\end{figure}   
From equations (3) and (4) we have\\
 $\frac{C^*_{LSA}(S_{1})}{C^*_{OPT}({S}_{1})} \leq   \frac{2m-2}{m} \leq (2-\frac{2}{m}) $. \textit{OPT} and \textit{LSA } perform equivalently for ${S}_1$ with $m=2$ as it is required to schedule only $2$ jobs. Therefore, it is proved that \textit{LSA} is ($2-\frac{2}{m}$)-competitive for ${S}_{1}$, where $m\geq3$.\\\\
\textbf{Theorem 3}. \textit{LSA is ($2-\frac{m^2-m+1}{m^2}$)-competitive for ${S}_{2}$, where $m\geq2$.}\\\\
\textit{Proof:} Let, $C^{*}_{OPT}({S}_{2})$ and $C^{*}_{LSA}({S}_{2})$ denote the makespan of \textit{OPT} and \textit{LSA} respectively for ${S}_2$. We ignore $T$, while scheduling each incoming job.\\\\
\textit{Computation of LSA:} The worst input job sequence for ${S}_{2}$ appears when the largest job available at the end of the input job sequence. Therefore, the sequence $\sigma_2= \left<J_1, J_2, .......J_{m^2-m}, J_{m^2-m+1}\right>$ holds the worst sequence for ${S}_2$ where the jobs from $J_1$ to $J_{m^2-m}$ are of size $1$ unit each and the ${J_{m^2-m+1}}^{th}$ job is the largest job with size $m^2$ unit. Initially at time $t=0$, \textit{LSA} assigns $m$ jobs on $m$ machines in one slot and finish them at $t=1$. Subsequently, $m(m-1)$ jobs are scheduled in $m-1$ slots and are completed at $t=m-1$. Now, at $t=m-1$, we are left with last two jobs of size $1$ unit and $m^2$ unit respectively and the load of each machine is $m-1$. So, the last job finishes at $t=m-1+m^2$. 
Therefore, we have\\
\hspace* {3.2cm} $C^*_{LSA}(S_2) \leq m-1+m^2$ \hspace*{4.8cm}(5)\\
%We present the Gantt chart of \textit{LSA} for ${S}_2$ in figure \ref{fig:ganttchartoflsacase2}.\\
%\begin{figure}[h]
%\centering
%\includegraphics[scale=0.56]{ganttchartoflsacase2.png}
%\caption{Gantt chart of LSA for class 2}
%\label{fig:ganttchartoflsacase2}  
%\end{figure}\\       
\textit{Computation of OPT:}  \textit{OPT} schedules the largest job first. So, at time $t=0$, the largest job  $J_{m^2-m+1}$ is assigned to $M_1$ along with $m-1$ jobs to remaining $m-1$ machines. In the same fashion, $m(m-1)$ jobs are completed at $t=m$ and the last job finishes at $t=m^2$.
Therefore, we have \\
\hspace* {3.2cm} $C^*_{OPT}(S_2) \geq m^2$\hspace*{5.9cm}(6)\\
From equations (5) and (6) we have \\
\hspace* {1.2cm} $\frac{C^*_{LSA}(S_2)}{C^*_{OPT}(S_2)} \leq 2-(\frac{m^2-m+1}{m^2})$. \\
As we are not considering the single machine case, so we have \textit{LSA} is ($2-\frac{m^2-m+1}{m^2}$)- competitive for ${S}_{2}$, where $m\geq2$.
\section{Conclusion and Future Scope}\label{sec:Conclusion and Scope of Future Work}
In this paper, we have presented an alternate proof for ($2-\frac{1}{m}$)- competitiveness for algorithm \textit{LSA} for independent jobs. We have studied and analyzed the performance of \textit{LSA} by characterizing the input sequences into two special classes. We have shown that \textit{LSA} is ($2-\frac{2}{m}$)-competitive for special class($S_1$) of input sequence, where we have considered $(m-1)^2 + 1$ jobs with  processing times such as $1$ unit and $m$ unit respectively. We have also shown that \textit{LSA} is ($2-\frac{m^2-m+1}{m^2}$)-competitive by considering another class($S_2$) of input sequence with $m(m-1)+ 1$ jobs of with sizes such as $1$ unit and  $m^2$ unit respectively. The competitive ratios achieved by \textit{LSA} for $S_1$ and $S_2$ input sequence with different number machines are shown in table \ref{tab: Competitive Ratio of LSA for special class of Inputs with Different Number of Machines}.  It can be observed from our analytical results that increase in number of machines does not help \textit{LSA} to minimize makespan for $S_1$. However, the performance of \textit{LSA}  can be improved substantially with the increase in number of machines for $S_2$.\paragraph{}
\textbf{Future Scope}. It can be realized that the order of availability of the jobs has strong influence on the performance of \textit{LSA}. However, the characterization of the input sequence with known total number of jobs and their processing time can help to improve the competitive ratio of \textit{LSA}. Through input characterization, theoretical input sequences can be mapped to the real-world input sequences. It will be interesting to evaluate the performance of well-known online scheduling algorithms for practical input sequences. The performance of well-known online scheduling algorithms can be improved with better competitive results based on the practical input sequences. 

%\\ It will be interesting to further characterize the input job sequences based on real world applications that may lead to the design of improved competitive online scheduling algorithm.
\begin{table}
\centering
\caption{Competitive Ratio of LSA for Different Number of Machines}
\begin{tabular}{|p{1.8cm}|p{2.5cm}|p{2.5cm}|}
\hline
\textbf{Number of Machines}  & \textbf{Competitive Ratio for Class-1} &  \textbf{Competitive Ratio for Class-2}\\
\hline 
2 &  1.0000 & 1.2500 \\
\hline
3 &  1.3333 & 1.2222 \\
\hline
4 &  1.5000 & 1.1875 \\
\hline
5 &  1.6000 & 1.1600 \\
\hline 
10 & 1.8000 & 1.0900 \\
\hline
50 & 1.9600 & 1.0196 \\
\hline
100 &  1.9800 & 1.0099 \\
\hline 
\end{tabular}
\label{tab: Competitive Ratio of LSA for special class of Inputs with Different Number of Machines}
\end{table}

%%%% This page is for instructions only, once the article is finalize please omit the below text before creating the final PDF


\begin{thebibliography}{}


\bibitem{Graham1966}
Graham R.L.(1966) "Bounds for certain multiprocessor anomalies". \textit{Bell System Technical Journal}, \textbf{45}:1563-1581.
\bibitem{Borodin1998}
Borodin A., El-Yaniv R.(1998) "Online computation and competitive analysis". \textit{Cambridge University Press}, Cambridge. 
\bibitem{Pruhs2004}
Pruhs K., Sgall J. and Torng E.(2004) "Online scheduling".\textit{ Handbook on scheduling: Algorithms, models and performance analysis, CRC Press}.
\bibitem{Robert1985}
Robert E. Tarjan and Sleator.(1985) "Amortized computational complexity". \textit{SIAM Journal on Algebric and Discrete Methods}, \textbf{6}(2):306-318.
\bibitem{Faigle1989}
Faigle U., Kern W. and Turan G.(1989) "On the performance of online algorithms for partition problems". \textit{Acta Cybernetica}, \textbf{9}:107-119.
\bibitem{Aspnes1993}
Aspnes J., Azar Y., Fiat A., Plotkin S. and Waarts O.(1993) "Online load balancing with applications to machine scheduling and virtual circuit routing". \textit{$25^{th}$ ACM STOC}: 623-631.  
\bibitem{Bartal1992}
Bartal Y., Fiat A. and Rabani Y.(1992) "Competitive algorithms for distributed data management". \textit{In Proceedings of $24^{th}$ Annual ACM symposium on Theory of Computing}:39-50.
\bibitem{Baeza-Yates1993}
Baeza-Yates R.A., Culberson J.C. and Rawlins G.J.E.(1993) "Searching in the plane". \textit{Information and Computation}, \textbf{106}: 234-252, 1993.
\bibitem{Garey1979}
Garey M.R. and Jhonson D.S.(1979) "Computers and Intractability: A Guide to the Theory of NP-Completeness". \textit{Freeman}.
\bibitem{Graham1979}
Graham R.L., Lawer E.L., Lenstra J.K. and Rinnooy kan A.H.(1979) "Optimization and ap-proximation in deterministic sequencing and scheduling : A Survey". \textit{Annals of Discrete Mathematics, Elsevier}, \textbf{5}: 287-326.
\bibitem{Chen1998}
Chen B., Potts C.N. and Woeginger G.J.(1998) "A review of Machine scheduling: Complexity, Algorithms and Approximability". \textit{Handbook of Combinatorial Optimization, Kluwer Academic Publishers}, \textbf{3}:21-169.
\bibitem{Sgall}
Sgall J.(1998) "Online scheduling: A survey". \textit{Lecture Notes in Computer science, Springer}, \textbf{1442}:196-231.
\bibitem{Albers2009}
Albers S.(2009) "Online scheduling:  Introduction to Scheduling", edited by Y. Robert and F. Vivien. \textit{CRC Press}: 57-84.
\bibitem{Galambos1993}
Galambos G. and  Woeginger G.J.(1993) "An online scheduling heuristic with better worst case ratio than Graham's list scheduling". \textit{SIAM Journal of Computing}, \textbf{22}(2):349-355.
\bibitem{Bartal1992}
Bartal Y., Fiat A., Karloff H. and Vohra R.(1992) "New algorithms for an ancient scheduling problem". \textit{In Proceedings of the $24^{th}$ ACM Symposium on the Theory of Computing}, Victoria, Canada:51-58.
\bibitem{Karger1996}
Karger D.R., Phillips S.J. and Torng E.(1996) "A better algorithm for an ancient scheduling problem". \textit{Journal of Algorithms}, \textbf{20}(19):400-430.
\bibitem{Bartal1994}
Y. Bartal Y., Karloff H. and Rabani Y.(1994) "A better lower bound for online scheduling". \textit{Information Processing Letters}, \textbf{50}:113-116.
\bibitem{Chen1994}
Chen B., Vliet A.V. and  Woeginger G.J. "New lower and upper bound for online scheduling". \textit{Operation Research Letters}, \textbf{16}:221-230.
\bibitem{Albers1999}
Albers S.(1999) "Better bounds for Online scheduling". \textit{SIAM Journal on Computing}, \textbf{29}:459-473. 
\bibitem{Fleischer2000}
Fleischer R. and Wahl M.(2000) "Online scheduling revisited". \textit{Journal of Scheduling}, \textbf{3}:343-353.
\bibitem{Rudin III2001}
Rudin III J.F.(2001) "Improved bounds for the online scheduling problem", \textit{Ph.D Thesis. The University of Texas at Dellas}.
\bibitem{J.F Rudin III}
Rudin III J.F. and Chandrasekaran R.(2003) "Improved bounds for the online scheduling problem". \textit{SIAM Journal of Computing}, \textbf{32}(3): 717-735.
\bibitem{Englert2008}
Englert M., Ozmen D. and Westermann M.(2008) "The power of reordering for online minimum makespan scheduling". \textit{In Proceedings $49^{th}$ Annual IEEE Symposium on Foundations of Computer Science}.
\end{thebibliography}
\end{document}